\documentclass[preprint,showpacs,preprintnumbers,nofootinbib,amsmath,amssymb,aps,prc,longbibliography]{revtex4-1}
\usepackage{graphicx}
\usepackage{dcolumn}
\usepackage{bm}
\begin{document}
\title{On the importance of using exact pairing in the study of pygmy dipole resonance}
\author{N. Dinh Dang$^{1}$}
  \email{dang@riken.jp}
\author{N. Quang Hung$^{2}$}
 \email{hung.nguyen@ttu.edu.vn}

 \affiliation{1) Theoretical Nuclear Physics Laboratory, RIKEN Nishina Center
for Accelerator-Based Science,
2-1 Hirosawa, Wako City, 351-0198 Saitama, Japan\\
and Institute for Nuclear Science and Technique, Hanoi, Vietnam\\
2) School of Engineering, TanTao University, TanTao University Avenue, TanDuc Ecity, Duc Hoa, Long An Province, Vietnam}

\date{\today}
\begin{abstract}
The strength functions of giant dipole resonance (GDR) in oxygen $^{18 - 24}$O, calcium $^{50 - 60}$Ca, and tin $^{120 - 130}$Sn isotopes are calculated within the phonon damping model under three approximations: without superfluid pairing, including BCS pairing, and exact pairing gaps. The analysis of the numerical results shows that exact pairing decreases the two-neutron separation energy in light nuclei, but increases it in heavy nuclei as compared to that obtained within the BCS theory. In neutron-rich medium and heavy nuclei, exact pairing significantly enhances the strength located at the low-energy tail of the GDR, which is usually associated with the pygmy dipole resonance. The line shape of the GDR changes significantly with increasing the neutron number within an isotopic chain if the model parameter is kept fixed at the value determined for the stable isotope.
\end{abstract}

\maketitle
\section{Introduction}
\label{Intro}
The enhancement of the $E1$ strength at low energy around the particle-emission threshold of neutron-rich nuclei has been
identified as the manifestation of the pygmy dipole resonance (PDR).  In the most intuitive and common interpretation, the PDR in a neutron-rich nucleus is represented as the oscillation of the weakly-bound neutron skin against the isospin-symmetric core of protons and neutrons. The PDR's nature and properties such as its collectivity, the dependence of its energy-weighted sum of strength (EWSS) have been the subjects of several experimental and theoretical studies in recent years. (See, e.g. Refs. \cite{P1,P2,P3,P4,P5,P6} and Refs. \cite{Sagawa,Vretenar,Ansari,Litvinova,Sarchi,Terazaki,Dang,PDM,Ebata,Lanza,Inakura,Gamba,Tsoneva}, for some experimental and theoretical works, respectively.) These studies reveal that the nature of the so-called PDR is still an issue open to debate. Is the PDR indeed a collective motion to deserve that name or merely composed of non-collective excitations, remnants at the GDR's low-energy tail like those observed in $^{18}$O?
The answers to these questions are currently far from being conclusive. The macroscopic picture, which represents the PDR as a collective oscillation of the neutron excess against the stable core is rather crude. In particular, when the PDR is strongly coupled to the GDR, the presentation of the PDR as a collective oscillation is in principle no longer valid~\cite{Gamba}. There are several microscopic calculations in literature showing that the PDR has a non-collective nature (See, e.g. Refs. \cite{Sarchi,Tsoneva}). 

One of the major issues in the theoretical study of the PDR in medium and heavy nuclei is the discrepancy in the predictions of different approaches regarding the strength and collectivity of the PDR. For example, while the relativistic random-phase approximation (RRPA) seems to predict a prominent peak identified as the collective PDR below 10 MeV in $^{120, 132}$Sn and $^{122}$Zr~\cite{Vretenar,Litvinova}, the results of calculations including monopole pairing within the quasiparticle RPA (QRPA) do not expose any collective states in the low-energy region of the $E1$ strength distribution for $^{120, 132}$Sn~\cite{Sarchi}. One of the possible sources of such discrepancy may well lie in superfluid pairing.

It is well-known that superfluid pairing plays a crucial role in open shell nuclei in the vicinity of the neutron drip line, where the neutron Fermi surface is located very close to the continuum. However all the theoretical calculations of the PDR so far either neglected pairing, such as the relativistic RPA, or adopted the mean-field pairing. The latter is taken into account within the Hartree-Fock-Bogolyubov (HFB)~\cite{Terazaki,Ebata}, Hartree-Fock (HF) + BCS formalisms, or coupling of QRPA particle-hole (ph) states to more complicate configurations like the 2p2h ones within the particle-phonon coupling scheme~\cite{Sarchi} or coupling to two, three phonon components~\cite{Tsoneva}. Given the progress in the exact solutions of the pairing problem in recent years~\cite{Volya}, it is highly desirable to see how exact pairing affects the PDR as compared to the predictions given by the approaches employing the conventional mean-field pairing gap.

The goal of the present paper is to study the effect of exact pairing on the PDR within the phonon damping model (PDM). The latter was extended to include the pairing gap and applied to calculate the PDR in oxygen and calcium isotopes in Ref. \cite{PDM}. Within this approach the GDR is generated by a phenomenological phonon with the unperturbed energy $\omega_q$. Because of coupling to ph configurations (for closed-shell nuclei) or 
two-quasiparticle configurations (for open-shell nuclei), this energy is shifted to the GDR energy $E_{GDR}$ and the GDR also acquires a width $\Gamma_{GDR}$. The PDR appears in this model as an enhancement of the strength at low energy (below 10 - 12 MeV) in the GDR strength function. By calculating the EWSS of the PDR and comparing it to the GDR energy-weighted sum rule, one can see the contribution of the PDR to the total GDR sum rule. To see the effect of exact pairing, apart from the BCS gap, which was used in the calculations in Ref. \cite{PDM}, the present paper also employs the exact gap, which is extracted from the exact pairing energy obtained by diagonalizing the pairing Hamiltonian.

The paper is organized as follows. The formalism is presented in Sec. \ref{formalism}. The results of numerical calculations for oxygen, calcium and tin isotopes within three schemes, namely without pairing, including BCS pairing, and exact pairing, are analyzed in Sec. \ref{results}. The paper is summarized in the last section, where conclusions are drawn.
\section{Formalism}
\label{formalism}
\subsection{Quasiparticle representation of the PDM}
\label{PDM}
We employ the quasiparticle representation of the PDM, discussed thoroughly in Ref. \cite{PDM}, where it was applied to study of the PDR in oxygen and calcium isotopes within the HF+BCS approach. Therefore, we do not repeat the detailed derivation already presented in Ref. \cite{PDM}, but summarize only the final expressions therein (for the zero temperature case), which are necessary for the calculations in the present paper. Instead of the HF single-particle energies, the present paper employs the single-particle spectra generated by the Woods-Saxon potentials for the isotopes under consideration.  

The PDM uses a model Hamiltonian, which consists of three terms. The first term describes the independent quasiparticle mean field with quasiparticle energies $E_j$, the second term stands for the phonon field with phonon energies 
$\omega_q$, whereas the last term treats the 
coupling between these two fields [See Eq. (1) in Ref. \cite{PDM}]. 
Because of this coupling, the GDR, which is generated by the phonon vibration, acquires a width $\Gamma_{GDR}$ and an energy shift from the unperturbed energy $\omega_q$ to the GDR energy $E_{GDR}$. 
The propagation of the phonon $q$ under this coupling is described by the Green function, whose final form is
\begin{equation}
G_{q} = \frac{1}{2\pi}\frac{1}{E-\omega_{q}-P_q(E)}~.
\label{G}
\end{equation}
The polarization operator $P_q(E)$ is given as
\begin{equation}
P_q(E) = \sum_{j \leq j'}[F_{jj'}^{(q)}u_{jj'}^{(+)}]^{2}\bigg[\frac{1}{\omega_q-E_j-E_{j'}}-\frac{1}{\omega_q+E_j+E_{j'}}\bigg],
\label{P}
\end{equation}
where $F_{jj'}^{(q)}$ are the matrix elements of the quasiparticle-phonon coupling, $u_{jj'}^{(+)}=u_jv_{j'}+v_ju_{j'}$ are the combinations of the 
$u_j$ and $v_j$ coefficients of the Bogolyubov's transformation from particles to quasiparticles, $E_j = \sqrt{(\epsilon_j-\lambda)^2+\Delta^2}$ are the quasiparticle energies, which are calculated from energies $\epsilon_j$ of single particles on the spherical orbitals $j$ with the chemical potential $\lambda$ and pairing gap $\Delta$. Notice that, for the GDR, when $\omega_q\simeq E_{GDR}$, the contribution of the 
second term on the right-hand side of Eq. (\ref{P}) is negligible compared to that of the first one.

The phonon damping $\gamma_q(\omega)$ at real $\omega$ is calculated as the imaginary part of the analytic continuation of the polarization operator $P_{q}(E)$ into the complex energy plane, that is $\gamma_q(\omega) = {\rm Im}[P_q(\omega\pm i\varepsilon)]$ with a sufficiently small value of the smoothing parameter $\varepsilon$. By using the $\delta$-function representation $\delta(x) = \varepsilon/[\pi(x^2+\varepsilon^2)]$, the final expression of $\gamma_q(\omega)$ is given as
\begin{equation}
\gamma_{q}(\omega) = \varepsilon\sum_{j \leq j'}[F_{jj'}^{(q)}u_{jj'}^{(+)}]^{2}\bigg[\frac{1}{(\omega_q-E_j-E_{j'})^2+\varepsilon^2}
-\frac{1}{(\omega_q+E_j+E_{j'})^2+\varepsilon^2}\bigg].
\label{gamma}
\end{equation}
The GDR strength function $S_q(\omega)$ is calculated by definition from the analytic property of the Green function (\ref{G}), namely 
$S_q(\omega)=i[G_q(\omega+i\varepsilon) - G_q(\omega-i\varepsilon)]$, which finally gives
\begin{equation}
S(\omega) =\frac{1}{\pi}\frac{\gamma_q(\omega)}{[\omega-E_{GDR}]^2+\gamma_q^2(\omega)}~,
\label{S}
\end{equation}
where the GDR energy $E_{GDR}$ is found as the solution of the equation
\begin{equation}
E_{GDR}-\omega_q-P_q(E_{GDR})= 0~,
\label{EGDR}
\end{equation}
whereas the GDR full width at half maximum is defined as
\begin{equation}
\Gamma_{GDR}= 2\gamma_q(E_{GDR})~.
\label{Gamma}
\end{equation}
From the above summary it is obvious that the mechanism leading to the damping width of the GDR in Eq. (\ref{Gamma}) is different from Landau damping. The latter is not a real damping, but  simply a deviation from the GDR energy centroid when the GDR is presented as a superposition of ph states. It exists even in the harmonic limit, e.g. the (Q)RPA, where the GDR never damps being a collection of harmonic oscillators. Within the PDM Landau damping would be the case if there were more than one GDR phonon $q$ and all the polarization operators $P_q(E)$ in Eq. (\ref{G}) were zero. This has never been considered within the PDM so far.

The quantity that measures the PDR strength relative to the total GDR energy-weighted sum of strength is given by the ratio
\begin{equation}
r = \sigma_{PDR}/\sigma_{GDR}~,\hspace{2mm} \sigma_{PDR} =\int_0^{E_{up}}S(\omega)\omega d\omega~,\hspace{2mm} \sigma_{GDR} =\int_0^{E_{max}}S(\omega)\omega d\omega~,
\label{r}
\end{equation} 
with the strength function $S(\omega)$ (\ref{S}). The values of $E_{max}$ is determined so that, within the interval $0\leq\omega\leq E_{max}$,  the integrated cross section $\sigma_{GDR}$ for stable isotopes exhausts the Thomas-Reich-Kuhn (TRK) sum rule equal to $TRK=60 NZ/A$. For $^{18}$O, $^{50}$Ca and $^{120}$Sn we have $E_{max}=$ 50, 40 and 30 MeV, respectively. These values are extended to all nuclei within the same isotopic chains. Once the strength function $S(\omega)$ is normalized to the TRK as $S'(\omega)\equiv S(\omega)\times TRK/\sigma_{GDR}$, the ratio (\ref{r}) becomes the same as $\sigma_{PDR}'/TRK$, where $\sigma_{PDR}'$ is the normalized cross section integrated up to $E_{up}$. The values $E_{up}=$ 15, 12, and 10 MeV are adopted in the calculations for oxygen, calcium, and tin isotopes, respectively. 
\subsection{Exact pairing gap}
\label{exact}
Exact pairing employed in the present paper is obtained by directly diagonalizing the pairing Hamiltonian~\cite{Volya}
\begin{equation}
H_P=\sum_{jm}\epsilon_{j}a_{jm}^{\dagger}a_{jm}-
    G\sum_{jj'}\sum_{mm'>0}(-)^{j+j'-m-m'}a_{jm}^{\dagger}a_{j-m}^{\dagger}a_{j'-m'}a_{j'm'}~, 
\label{Hpair}
\end{equation}
which describes a system of $N$ particles interacting via a
    monopole-pairing force with a constant parameter $G$ and having single-particle energies
    $\epsilon_{j}$, generated by particle creation operators
    $a_{jm}^{\dagger}$ on the spherical $j$-th orbitals with shell degeneracies
    $2j+1$. 
    
This approach does not produce a pairing gap by itself, because the latter is a mean field concept. To 
mimic the mean-field pairing gap a pairing gap is introduced as [Cf. Eq. (18) of Ref. \cite{CE} and the discussion therein.]:
\begin{equation}
    \Delta_{ex}=\sqrt{-G{\cal E}_{\rm pair}}~,\hspace{5mm}
    {\cal E}_{\rm pair}={\cal E}-{\cal E}^{(0)}~, \hspace{5mm} {\cal
    E}^{(0)}\equiv
    \sum_{j}(2j+1)(\epsilon_{j}- \frac{G}{2}f_{j}^{ex})f_{j}^{ex}~,
    \label{gap}
    \end{equation}    
where $f_j^{ex}$ are the exact single-particle occupation numbers [Eq. (4) of Ref. \cite{CE}].  
Being constructed based on the exact eigenvalues of the paring problem, the gap (\ref{gap}) is referred to as the exact pairing gap hereafter. 

The exact chemical potential is calculated from the definition
\begin{equation}
    \lambda_{ex} = \frac{1}{4}[{\cal E}(N+2) - {\cal E}(N-2)]~,
    \label{lambda}
    \end{equation}
    where ${\cal E}(n)$ is the exact (ground-state) energy of the system with $n$ particles.
Given the exact chemical potential $\lambda_{ex}$ and pairing gap $\Delta_{ex}$, we calculate the exact quasiparticle energies $E_j^{ex}$ 
as 
\begin{equation}
E_j^{ex}=\sqrt{(\epsilon_j-\lambda_{ex})^2+\Delta_{ex}^2}~.
\label{qpenergy}
\end{equation}
Within the same approximation, the coefficients $u_j$ and $v_j$ are replaced with their exact counterparts, $\sqrt{1-f_j^{ex}}$ and $\sqrt{f_j^{ex}}$, respectively, based on the equation for the particle number $N$
\begin{equation}
N = \sum_{j}(2j+1)f_j^{ex} = \sum_j(2j+1)v_j^2~.
\label{N}
\end{equation}
With these exact pairing gap, quasiparticle energies, chemical potential, and the counterparts $\sqrt{1-f_j^{ex}}$ and $\sqrt{f_j^{ex}}$ of the coefficients $u_j$ and $v_j$, we can easily calculate the
GDR strength function (\ref{S}) with exact pairing.

Within the BCS theory, a quasiparticle is obtained by dressing the single-particle excitation with the pairing correlation, following the Bogoliubov's canonical transformation from single-particle operators to quasiparticle ones. In other words, a quasiparticle with the quasiparticle energy $E_j$ is a Cooper pair of two time-conjugated particles with occupation numbers $u_j$ and $v_j$, respectively. 
The quasiparticle excitation with energies $E_j^{ex}$ (\ref{qpenergy}) and occupation numbers $\sqrt{1-f_j^{ex}}$ and $\sqrt{f_j^{ex}}$ should be understood as a modification or improvement of the conventional BCS quasiparticle excitation.
The justification for such an approximation can be seen in the cases where both the exact and the particle-number projected BCS gaps are possible (including the corrections owing to coupling to the self-consistent QRPA). By using the same value of the pairing interaction parameter $G$ and the same single-particle space, these pairing gaps are rather close to each other and behave similarly as functions of excitation energy (or temperature), especially at $T\leq$ 2 MeV. (See Fig. 1 of Ref. \cite{CE}.) 
\section{Analysis of numerical results}
\label{results}
\subsection{Ingredients of the numerical calculations}
    \begin{figure}
       \includegraphics[width=16cm]{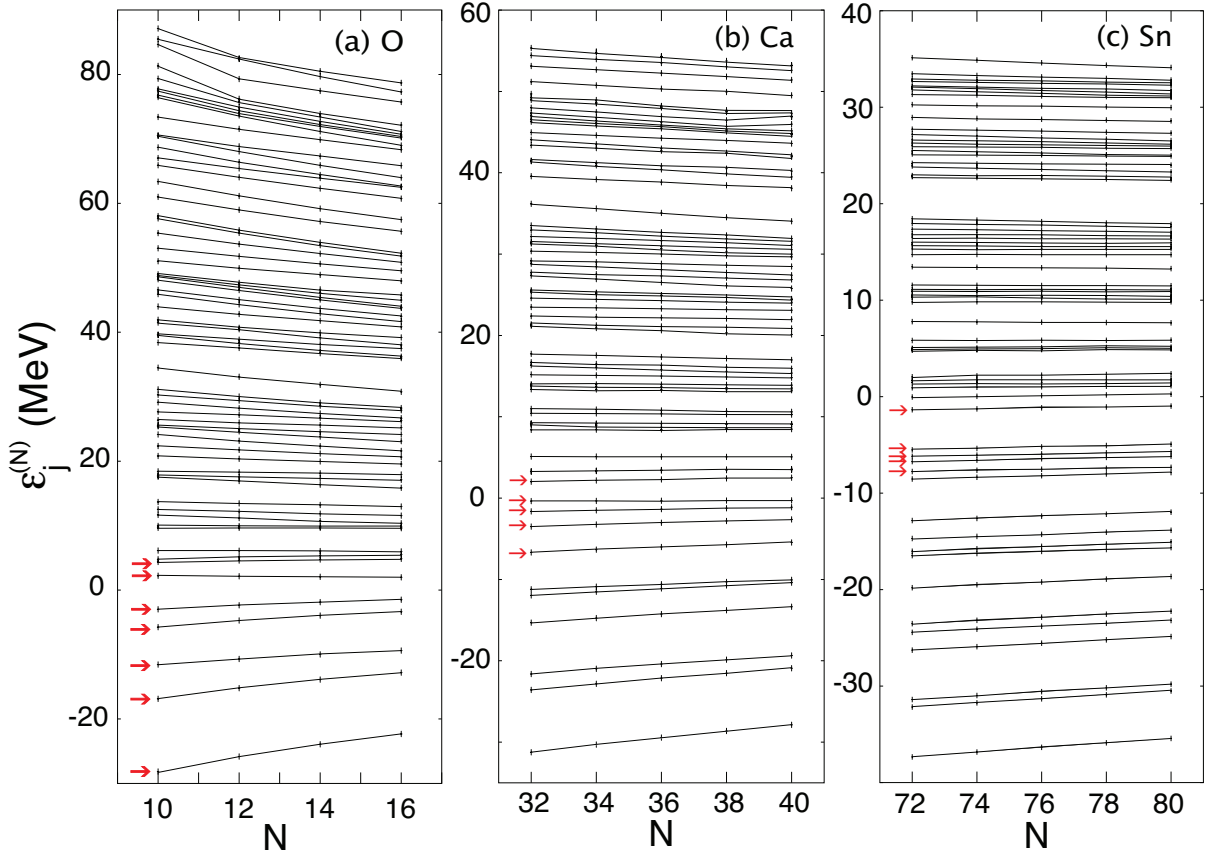}
        \caption{(Color online) Neutron single-particle energies $\epsilon_j^{(N)}$ versus the neutron number $N$ for oxygen (a), calcium (b) and tin (c) isotopes under consideration. The (red) arrows show the levels used in the diagonalization of the pairing Hamiltonian (See text). The lines connecting 
        the discrete energies at different $N$ are drawn as a guide to the eye.
        \label{spectra}}
    \end{figure}
The neutron and proton single-particle energies for $^{18 - 24}$O, $^{50 - 60}$Ca, and $^{120 - 130}$Sn are obtained within the Woods-Saxon potentials, which include the spin-orbit and Coulomb interactions. The depth of the central potential is given as
$V = V_0 \left[1 \pm k (N-Z)/(N+Z)\right]$ with $V_0=$ 51 MeV and $k=$ 0.86, whereas the plus and minus signs stand
for proton ($Z$) and neutron ($N$), respectively. The radius $r_0$,
diffuseness $a$, and spin-orbit strength $\lambda$ are chosen to be $r_0
=$ 1.27 fm, $a =$ 0.67 fm and $\lambda =$ 35.0, which are close to the parametrazation by Blomqvist and Wahlborn and the universal parametrization (See Tale 1 of Ref. \cite{WS}). The single-particle energies span a large space from the bottom level $1s_{1/2}$ up to around 80 - 87 MeV for oxygen isotopes, 50 - 60 MeV for calcium isotopes, and 35 - 40 MeV for tin isotopes, 
from which the high-lying positive energy levels form a kind of discretized continuum. Our experience shows that such large configuration spaces
are sufficient to describe the GDR and PDR~\cite{Dang,PDM,PDM2}. The neutron single-particle energies employed in this paper are plotted against the neutron number in Fig. \ref{spectra} for all isotopic chains under consideration. An increase of the level density with $N$ is clearly seen, which is stronger for oxygens, and much weaker for tins.

    \begin{figure}
       \includegraphics[width=16cm]{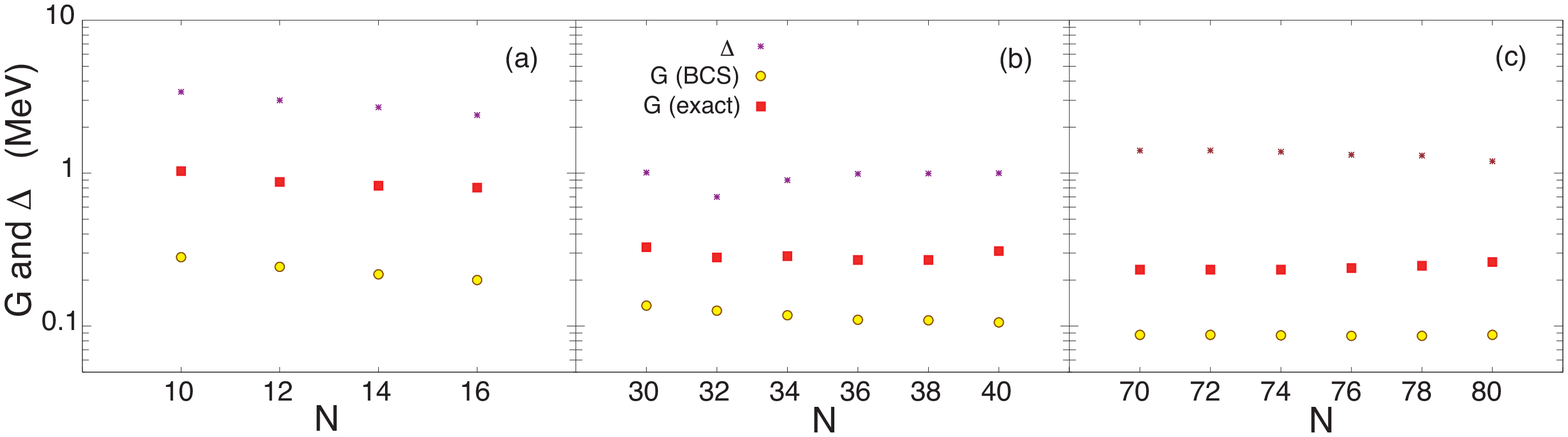}
        \caption{(Color online) Values of pairing parameter $G$ used in the BCS (yellow circles) and exact (red boxes) calculations as well as of the calculated pairing gap (purple crosses) for neutrons versus the neutron number $N$ for oxygen (a), calcium (b), and tin (c) isotopes.
        \label{Ggap}}
    \end{figure}
The values of pairing parameter $G$ for neutrons are chosen so that the pairing gaps obtained within the BCS and by exact diagonalization of the pairing Hamiltonian (\ref{Hpair}) are within the range of experimental values of the three-point gaps in oxygen, calcium and tin isotopes under consideration~\cite{Bohr,exgap}. Because of the limitation on the size of the matrix to be diagonalized, only single-particle orbitals around the Fermi level for neutrons are selected
in the exact diagonalization of the pairing Hamiltonian (\ref{Hpair}). For oxygens isotopes, seven $(2j+1)$-degenerate spherical orbitals, starting from the bottom one, are selected, namely $1s_{1/2}1p_{3/2}1p_{1/2}1d_{5/2}2s_{1/2}1d_{3/2}2p_{3/2}$. For calcium isotopes, five spherical orbitals are used, namely $1f_{7/2}2p_{3/2}1f_{5/2}2p_{1/2}1g_{9/2}$. For tin isotopes, the diagonalization includes five spherical orbitals $1g_{7/2}3s_{1/2}2d_{3/2}1h_{11/2}2f_{7/2}$, where for the uppermost one of this set, $2f_{7/2}$, two, three, and four doubly-degenerate deformed levels are taken for $^{120-126}$Sn, $^{128}$Sn, and $^{130}$Sn, respectively. All the levels within the selected sets are assumed to have the same pairing gap $\Delta_{ex}$ and the corresponding exact single-occupation numbers $f_{j}^{ex}$. The orbitals beyond these selected sets are assumed to have no pairing, i.e. $f_{j}^{ex}=$ 1 and 0 for $j$-th orbital located below and above the Fermi level, respectively. As for the BCS calculation of the pairing gap, the above-mentioned entire single-particle spectra are used 
because of two reasons. First, no limitation is required on the configuration space within the BCS theory. Second, the BCS gap obtained in the full single-particle spectra is the one, which is often used in various calculations of the PDR that include pairing correlations. 
The values of $G$ employed in the BCS and in the diagonalization of the pairing Hamiltonian (\ref{Hpair}) as well as the neutron pairing gaps themselves are plotted against the neutron number $N$ of the isotopes under study in Fig. \ref{Ggap}.
The values for $G$ used in the BCS calculation are naturally much smaller than those used in diagonalizing the pairing Hamilatonian (\ref{Hpair}) to obtain the exact pairing gap because the BCS calculation is extended to the entire single-particle spectrum to give the same pairing gap obtained within a much smaller configuration space in the diagonalization. 

    \begin{figure}
       \includegraphics[width=8.5cm]{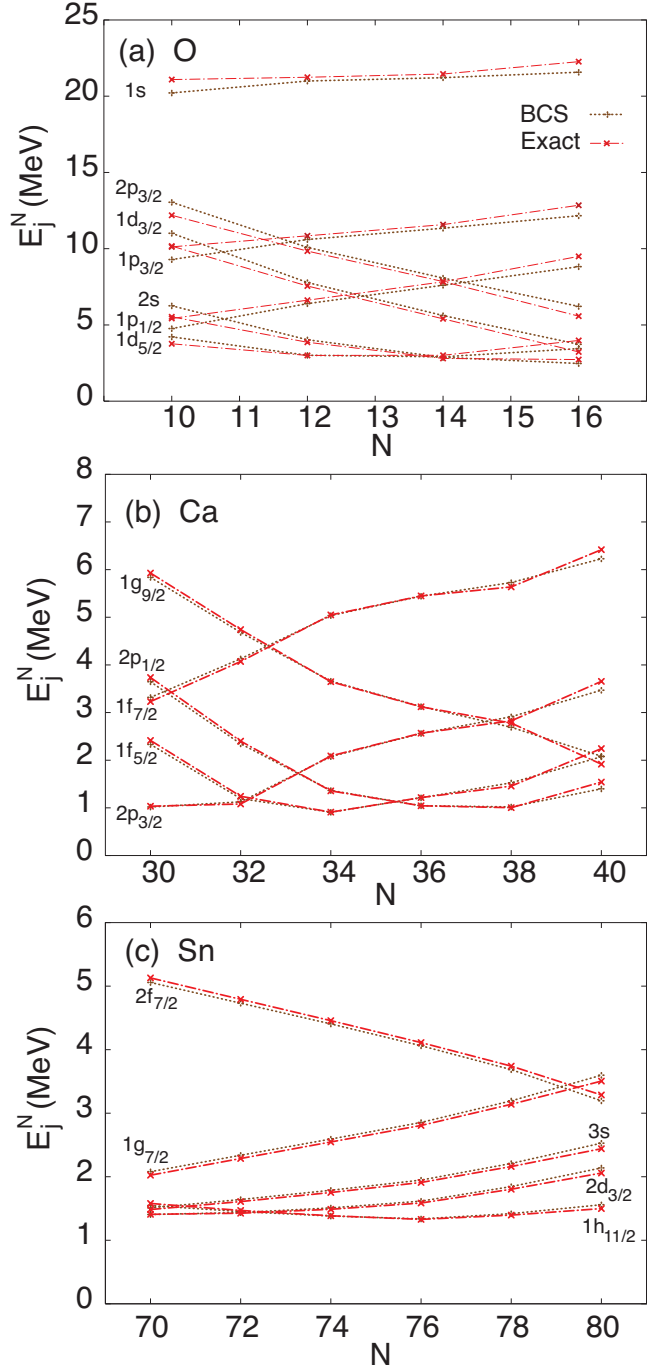}
        \caption{(Color online) BCS ((brown) crosses) and exact ((red) saltires) neutron quasiparticle energies $E_j^{N}$ versus the neutron number $N$ for the selected orbitals employed in the diagonalization of the pairing Hamiltonian (see text) in oxygen (a), calcium (b), and tin (c) isotopes. The (brown) dotted and (red) dash-dotted lines connecting the discrete energies at different $N$ are drawn as a guide to the eye.
        \label{Ejrev}}
    \end{figure}
\begin{figure}
       \includegraphics[width=8.5cm]{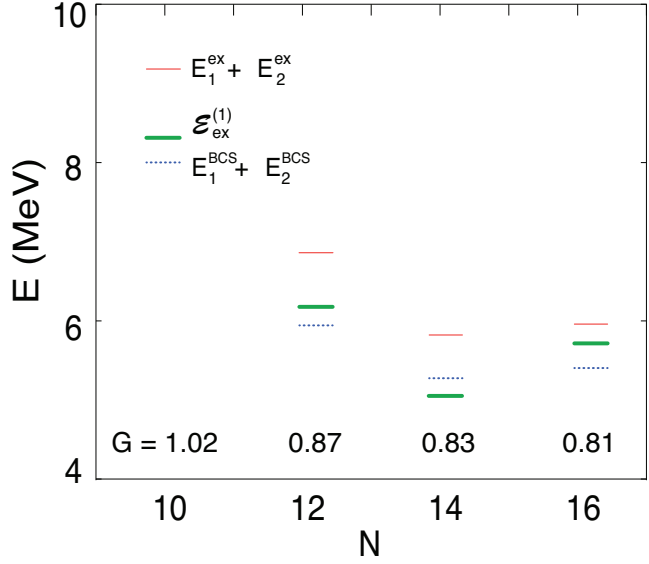}
        \caption{(Color online) The BCS and exact energies $E_1+E_2$ of the lowest two-quasiparticle state  as compared with the difference ${\cal E}_1^{ex}$ between the exact energy of the first excited state and that of the ground state for neutrons in oxygen isotopes. The values (in MeV) of the pairing interaction parameter $G$ to obtain these results are shown at the bottom of the figure. 
        \label{Eex1rev}}
    \end{figure}

The BCS $E_j$ and exact quasiparticles energies $E_j^{ex}$ for the selected neutron orbitals employed in the diagonalization of the pairing Hamiltonian are plotted against the neutron number $N$ for oxygen, calcium and tin isotopic chains in Fig. \ref{Ejrev}. The effect of exact pairing on the quasiparticle energies is noticeable for oxygen isotopes, whereas it is quite weak for calcium and tin isotopes. A comparison between the BCS and the exact two-quasiparticle energies $E_1 + E_2$ of two lowest quasiparticle states ($|1\rangle$, $|2\rangle$)  with the difference ${\cal E}_1^{ex}$ between the exact energy of the first excited state and that of the ground state for neutrons in oxygen isotopes in presented Fig. \ref{Eex1rev}. The selected two neutron orbitals are those located closest to the Fermi level, namely $(1d_{5/2}, 1p_{1/2})$ for $^{18}$O, $(1d_{5/2}, 2s_{1/2})$ for $^{20,22}$O, and $(2s_{1/2}, 1d_{3/2})$ for $^{24}$O. The BCS results shown in this figure are obtained by using the same value of the pairing constant $G$ and the same single-particle space that produce the exact pairing gap $\Delta_{ex}$ in Fig. \ref{Ggap} (a).  From Fig. \ref{Eex1rev} it is seen that the exact two-quasiparticle energy $E_1^{ex} + E_2^{ex}$ is always larger than ${\cal E}_1^{ex}$, whereas $E_1^{BCS} + E_2^{BCS} > {\cal E}_1^{ex}$ only in one case with $N=14$. This feature shows that the effect of  exact pairing shifts the two-quasiparticle energy toward the direction of the solution of the QRPA, where the energy of the first excited state is always larger than the exact one. (See the (green) dash-dotted line in Fig. 5 in Ref. \cite{PRC76}.)

The PDM assumes that the matrix elements $F_{jj'}^{(q)}$ of GDR coupling to ph configurations, causing the quantal width, are all equal to ${F}_1$. [See Sec. II B of Ref. \cite{PDM2} for the detail discussion on the justification of this assumption.] Because the microscopic origin of the spreading width $\Gamma^{\downarrow}$ at zero temperature ($T=$ 0) comes from coupling of 1p1h to more complicate configurations such as the 2p2h ones, whereas the PDM phonon has no 1p1h structure,  the parameter ${F}_1$ is generally selected to reproduce the GDR experimental width at $T=$ 0, which is essentially the sum of the spreading width $\Gamma^{\downarrow}$ and the escape width $\Gamma^{\uparrow}$. The latter comes from coupling to continuum and is in the order of few hundred keVs. It is effectively included via the smoothing parameter $\varepsilon$ in Eq. (\ref{gamma}) within the PDM. This smoothing parameter is usually taken equal to 0.5 MeV. Applying the above-mentioned principle of selecting $F_1$, we choose the two schemes of parameter selection as follows:

{\it I) Fixed $F_1$:}

For each isotope chain the parameter $F_1$ is chosen to reproduce the experimental GDR width in the stable isotope, namely $^{16}$O for oxygen, $^{48}$Ca for calcium, and $^{120}$Sn for tin isotopes. For oxygen and calcium isotopes the corresponding values of $F_1$ are used to calculate the GDR without pairing for all nuclei within the same isotopic chain. Upon including the BCS and exact pairing gaps, the value of $F_1$ is readjusted to $F_1^{BCS}$ and $F_1^{ex}$, respectively, to produce the same GDR width as that obtained without pairing for $^{18}$O and $^{50}$Ca, separately.  For $^{120}$Sn the values $F_1$, $F_1^{BCS}$, and $F_1^{ex}$ are chosen so that the corresponding calculations, namely without pairing, including BCS, and exact pairing, produce the same GDR width of 4.9 MeV. 
These values $F_1$, $F_1^{BCS}$, and $F_1^{ex}$ are then used in calculations for all other tin isotopes. This scheme is referred to as ``parameter selection I" hereafter.

{\it II) Width fitting:}

In this scheme, for each isotope the values $F_1^{BCS}$ and $F_1^{ex}$ are readjusted to obtain the same GDR width, which is given by the calculations using the value $F_1$ (without pairing) in the same isotope. We refer to this scheme as ``parameter selection II".

Regarding the GDR energy, the solution of Eq. (\ref{EGDR}) is usually required in calculations of the GDR width at finite temperature ($T\neq$ 0)~\cite{PDM2,PDMT} to ensure that, by choosing an appropriate value of the parameter ${F}_2 = F_{pp'}^{(q)} = F_{hh'}^{(q)}$ of coupling to pp and hh configurations at $T=$ 0, the solution of Eq. (\ref{EGDR}) does not deviate strongly from the GDR energy $E_{GDR}$ as $T$ varies,  in agreement with the experimental systematics for GDR energy in hot nuclei. The present work considers only the zero temperature case ($T=$ 0), where the term containing the parameter $F_2$ vanishes because $n_p - n_{p'} = n_h - n_{h'} =$0 (see the discussion after Eq. (15) in Ref. ~\cite{PDM}). So there remains only coupling of GDR phonon to particle-hole configurations, that is parameter $F_1$.  Therefore, one can either choose the unperturbed phonon energy $\omega_q$  close to $E_{GDR}$ and solve Eq. (\ref{EGDR}) to obtain the solution equal to $E_{GDR}$, or simply use the energy $E_{GDR}$ directly in Eq. (\ref{S}) without the need of choosing $\omega_q$ and solving Eq. (\ref{EGDR}). Alternatively, if one fixes $\omega_q$ at the value chosen for $^{18}$O e.g., and uses it for all other oxygen isotopes to solve Eq. (\ref{EGDR}), as has been done in Ref. \cite{PDM}, the GDR energy $E_{GDR}$ will change slightly with A as shown in Table I of Ref. \cite{PDM}. This does not change the GDR width (\ref{Gamma}) as it depends neither on $\omega_q$ nor $E_{GDR}$. Because not every experimental value of $E_{GDR}$ for neutron-rich isotopes is known and because a shift by around 1 MeV in the GDR energy $E_{GDR}$ does not really affect the PDR, we avoid choosing $\omega_q$ and solving Eq. (\ref{EGDR}) in the present paper for simplicity, and assign $E_{GDR}$ for oxygen isotopes to be equal to the experimental  GDR energy in $^{18}$O, which is 23.7 MeV~\cite{O18}. For calcium and tin isotopes the empirical fitting $E_{GDR}=31.2 A^{-1/3} + 20.6 A^{-1/6}$ for medium and heavy nuclei is adopted~\cite{Berman}. 

    \begin{figure}
       \includegraphics[width=12cm]{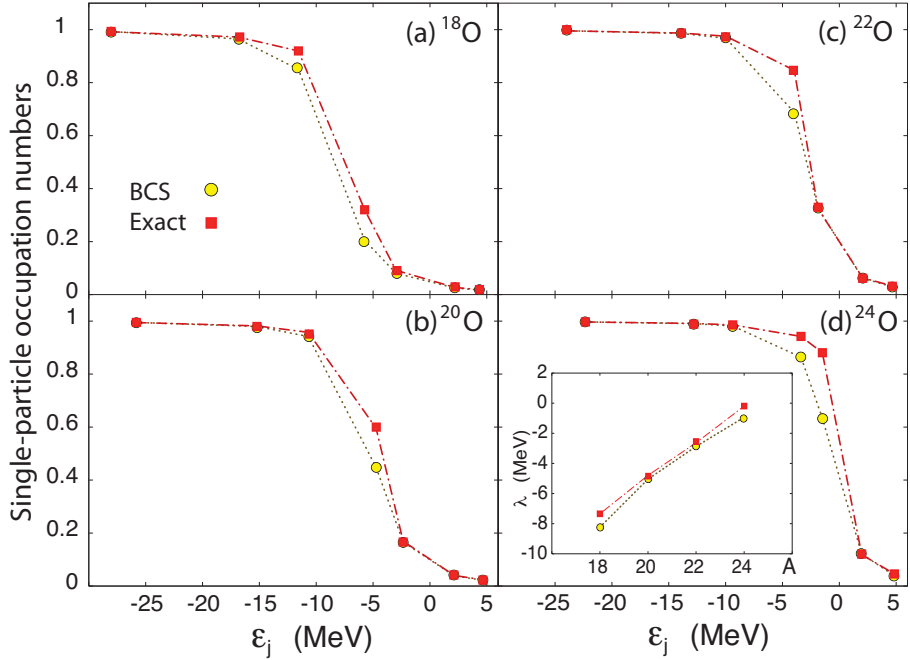}
       \caption{(Color line) Neutron single-particle occupation numbers $f_j^{ex}$ (red boxes), predicted by the diagonalization of the pairing Hamiltonian (\ref{Hpair}), and $v_j^{2}$ (yellow circles), obtained within the BCS theory vs the corresponding single-particle energies $\epsilon_j$ for oxygen isotopes. The inset shows the neutron chemical potentials obtained within the exact and BCS calculations as functions of the mass number. The lines are drawn as a guide to the eye.
        \label{v2O}}
    \end{figure}
    \begin{figure}
       \includegraphics[width=12cm]{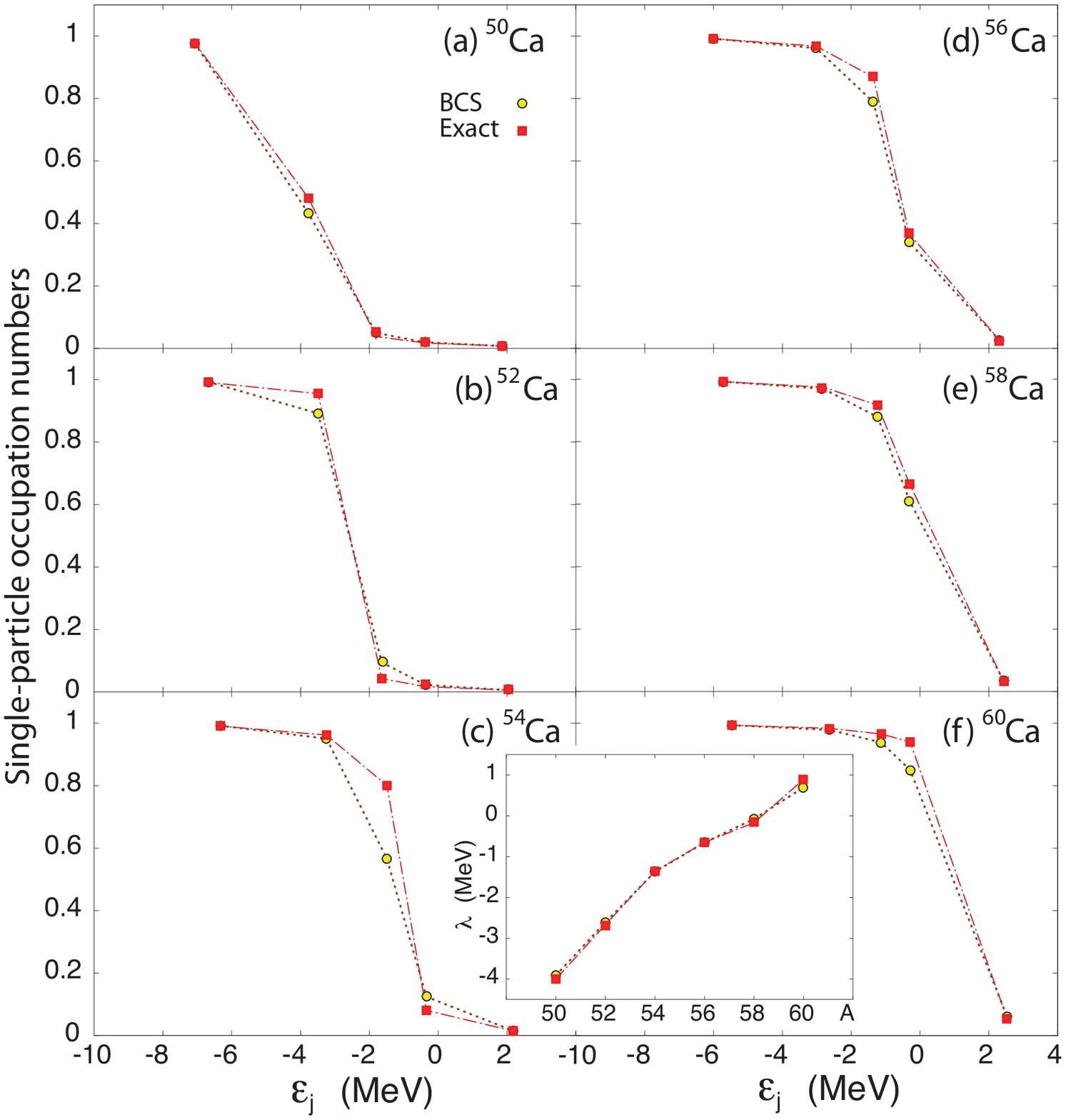}
       \caption{(Color online) Same as Fig. \ref{v2O} for calcium isotopes.
        \label{v2Ca}}
    \end{figure}
    \begin{figure}
       \includegraphics[width=12cm]{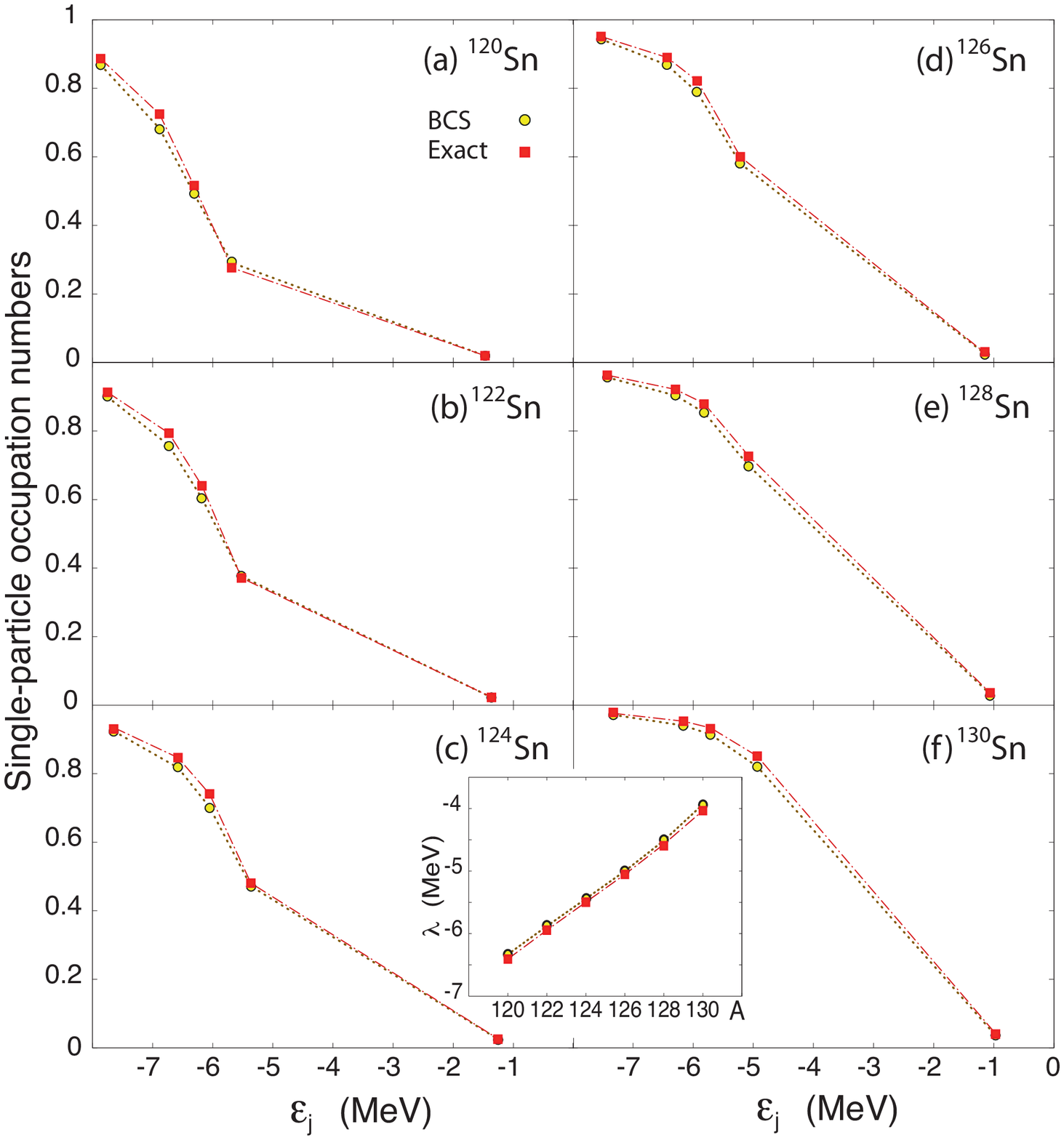}
       \caption{(Color online) Same as Fig. \ref{v2O} for tin isotopes.
                       \label{v2Sn}}
    \end{figure}
\subsection{Neutron single-particle occupation numbers and chemical potentials} 
The neutron single-particle occupation numbers predicted by the diagonalization of the pairing Hamiltonian (\ref{Hpair}) and BCS calculations as $f_j^{ex}$ and $v_j^2$, respectively, are shown in Fig. \ref{v2O} for oxygens, Fig. \ref{v2Ca} for calciums, and Fig. \ref{v2Sn} for tins against their corresponding single-particle energies within the selected single-particle sets. These figures show that the exact chemical potential is slightly larger than its value predicted by the BCS for light nuclei (oxygens), almost the same as the BCS value for medium nuclei (calciums), and becomes slightly smaller than the BCS value in heavy nuclei (tins). In other words, exact pairing decreases the two-neutron separation energy in light nuclei, but increases it in heavy nuclei, as compared to the BCS one. The difference between the effect of exact and BCS pairings on the occupations numbers is stronger in light nuclei (oxygens), whereas in heavy nuclei (tins) it is rather small. In all the cases considered here,  $f_j^{ex}$ is always larger than $v_j^{2}$  for the level closest to the Fermi surface. They also demonstrate that the single-particle levels selected for the diagonalization of the pairing Hamiltonian (\ref{Hpair}) are reasonable because, beyond them, $v_j^{2}$ are very close or practically equal to 1 or 0 depending on whether the level is below or above the Fermi surface.  The only exception is the orbital $2d_{5/2}$ below -8 MeV in tins, for which $v_j^2\simeq$  0.93 and 0.94 for $^{120}$Sn and $^{122}$Sn, respectively. This systematics turns out to give a significant effect on the PDR as will be seen in the next section.

\begin{figure}
       \includegraphics[width=12cm]{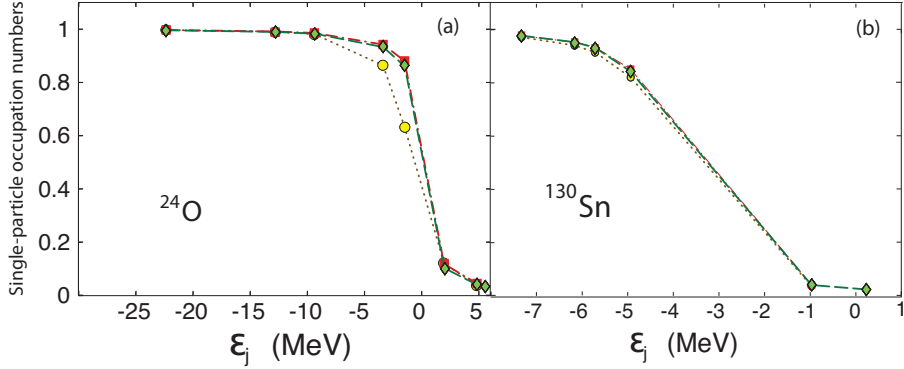}
       \caption{(Color line) Same as Figs. \ref{v2O} (d) and \ref{v2Sn} (f) for $^{24}$O and $^{130}$Sn, respectively. The (green) diamonds connected with dashed lines are exact occupation numbers $f_j^{j}$ obtained by enlarging the single-particle space used in diagonalizing the pairing Hamiltonian (see text).
                      \label{test}}
    \end{figure}

To double check that the selected single-particle levels are sufficient for exact pairing, a test is conducted for the two most neutron-rich nuclei, from the lightest and heaviest isotopes under consideration, namely $^{24}$O and $^{130}$Sn, respectively. For the neutron spectrum in $^{24}$O, three doubly degenerated levels from the orbital  
1$g_{7/2}$ at energy around 5.51 MeV are added on top of the orbit 2$p_{3/2}$ at 4.84 MeV in diagonalizing the pairing Hamiltonian (\ref{Hpair}). For the neutron spectrum in $^{130}$Sn, one double degenerated level from the orbital 3$p_{3/2}$ at energy around 0.26 MeV is added on top of the orbital 2$f_{7/2}$. The pairing parameter $G$ is readjusted to reproduce the same values for the pairing gaps in these nuclei used before this spectrum enlargement. As a result, the value of $G$ reduces from 0.811 MeV to 0.672 MeV for $^{24}$O and from  0.261 MeV to 0.256 MeV for $^{130}$Sn. The exact occupation numbers $f_j^{ex}$ obtained in this test are shown in Fig. \ref{test} as the (green) diamonds connected with dashed lines, which practically coincide with the results before the spectrum enlargement (red boxes with dot-dashed lines). This is an additional confirmation that the selected single-particle levels are sufficient to obtain the adequate exact pairing for the isotopes under consideration. 

\subsection{E1 strength function} 
    \begin{figure}
       \includegraphics[width=16cm]{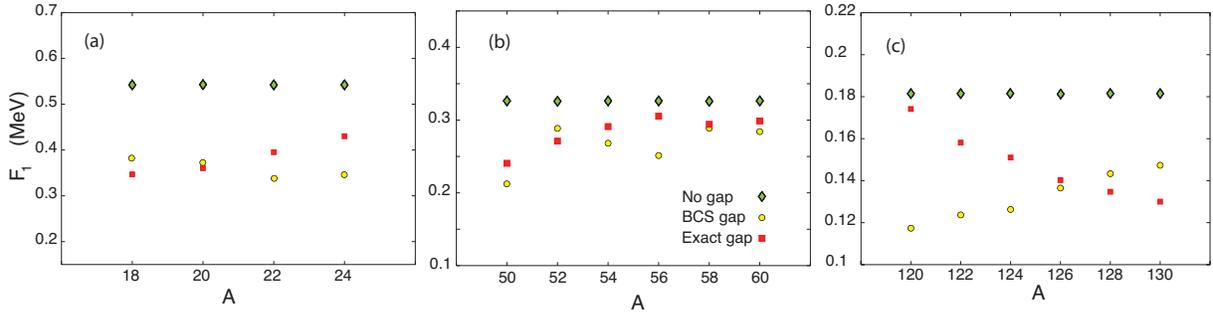}
        \caption{(Color online) Values of the particle-phonon coupling parameter $F_1$ within the parameter selection II adopted in the PDM calculations without pairing (green diamonds), including BCS pairing (yellow circles), and exact pairing (red boxes) for oxygen (a), calcium (b), and tin (c) isotopes.         \label{F1}}
    \end{figure}
The values $F_1$, $F_1^{BCS}$, and $F_1^{ex}$ within parameter selection II are shown in Fig. \ref{F1} as functions of the mass number $A$. The parameter selection I corresponds to the values of $F_1$ at $A=$ 18, 50, and 120 for oxygen, calcium, and tin isotopes, respectively. This figure shows a striking difference between the values of $F_1$ in the non-pairing, BCS-pairing, and exact-pairing cases.  This difference implies that, in general, by including pairing interaction, one should use weaker coupling between the GDR phonon and quasiparticle pairs to obtain the same GDR width as that predicted by the calculations without pairing. The dependences of $F_1^{BCS}$ and $F_1^{ex}$ on $A$ have opposite trends for light and heavy isotopes [Figs. \ref{F1} (a) and \ref{F1} (c)], namely $F_1^{ex}$ increases for oxygens but decreases for tins with increasing $A$, whereas for $F_1^{BCS}$ a reverse trend is seen.
The intermediate stage takes place for calcium isotopes [Figs. \ref{F1} (b)], where $F_1^{ex}$ first increases with $A$ but then remains at approximately the same value at $A\geq$ 54, whereas $F_1^{BCS}$ shows some oscillating behavior as a function of $A$. For neutron-rich medium-mass isotopes, namely $^{52 - 60}$Ca, the values $F_1^{ex}$ and $F_1^{BCS}$ are not much different from $F_1$, whereas for neutron-rich oxygen and tin isotopes they are significantly smaller than $F_1$. These results confirm the observation in microscopic calculations using the effective interactions, such as various types of the Skyrme force, that one cannot adopt the same parameter set fixed for stable nuclei, at least  in calculations of GDR for nuclei close to the neutron drip line.  
\subsubsection{Predictions by using parameter selection I}
    \begin{figure}
       \includegraphics[width=11cm]{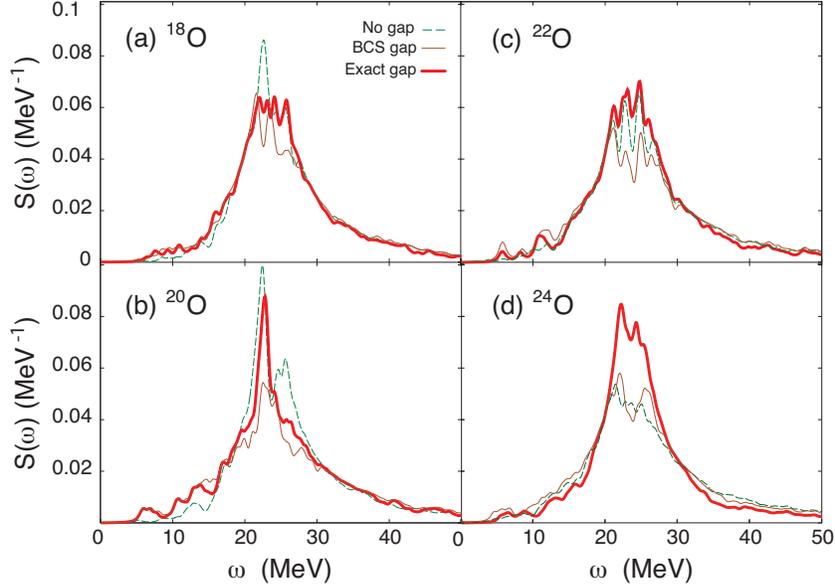}
        \caption{(Color online) GDR strength functions for oxygens isotopes obtained by using parameter selection I.  The predictions without pairing, including BCS pairing and exact pairing are denoted by the (green) dashed, (brown) thin solid, and (red) thick solid lines, respectively. \label{sOF1}}
    \end{figure}
    \begin{figure}
       \includegraphics[width=11cm]{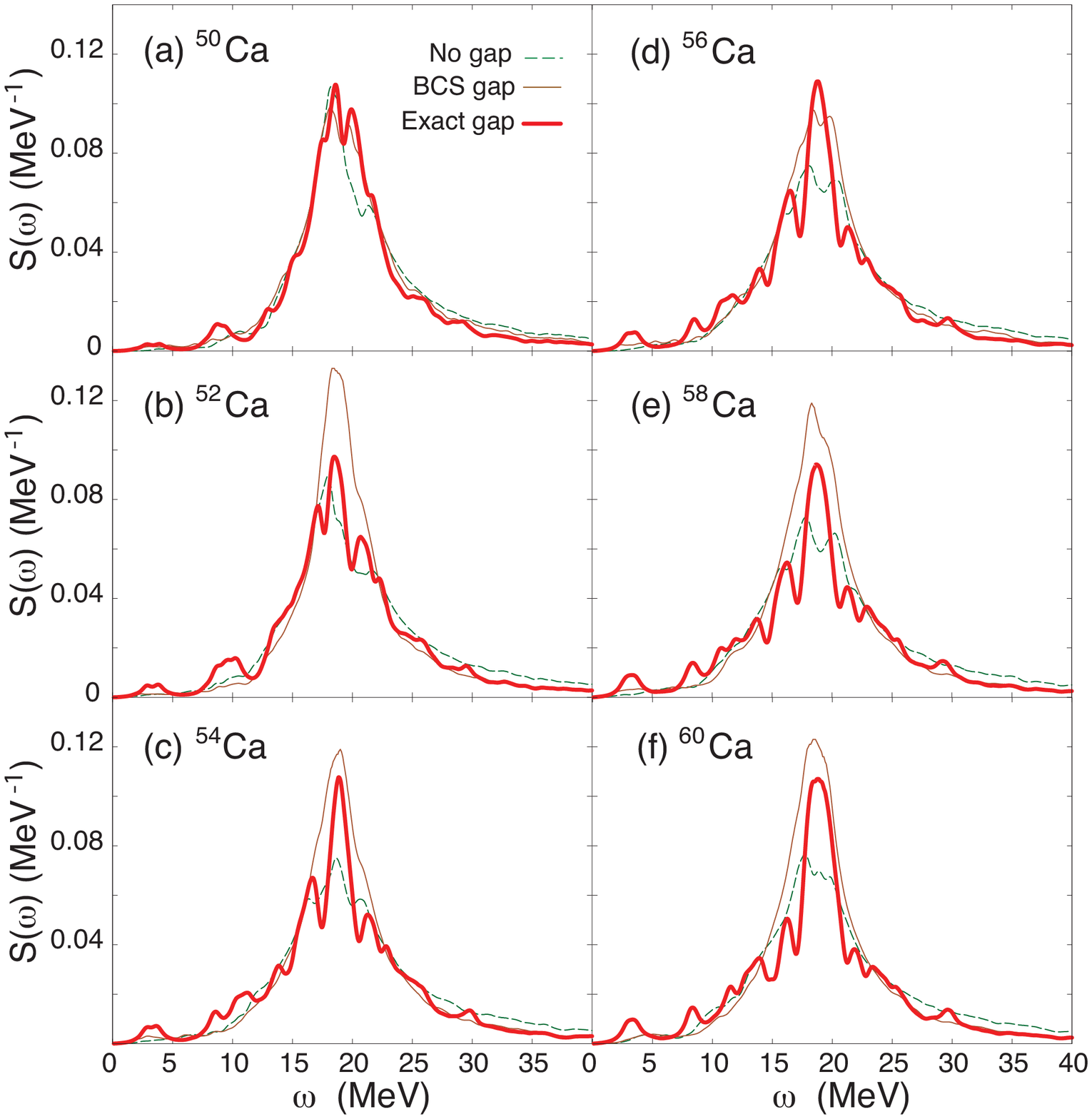} 
       \caption{(Color online) Same as Fig. \ref{sOF1} for calcium isotopes.
         \label{sCaF1}}
    \end{figure}
    \begin{figure}
       \includegraphics[width=11cm]{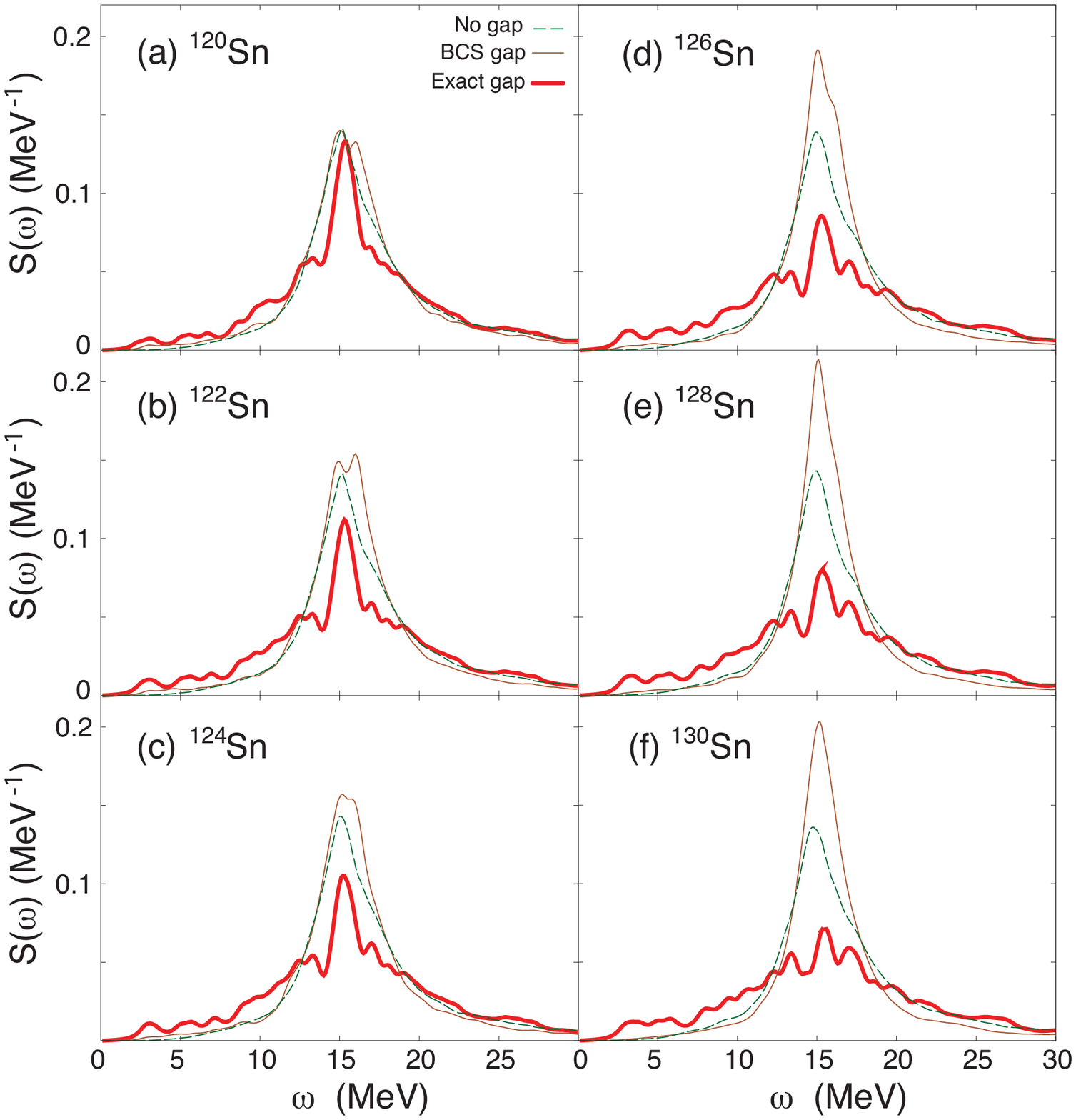}
        \caption{(Color online) Same as Fig. \ref{sOF1} for tin isotopes.
        \label{sSnF1}}
    \end{figure}
\begin{figure}
       \includegraphics[width=16cm]{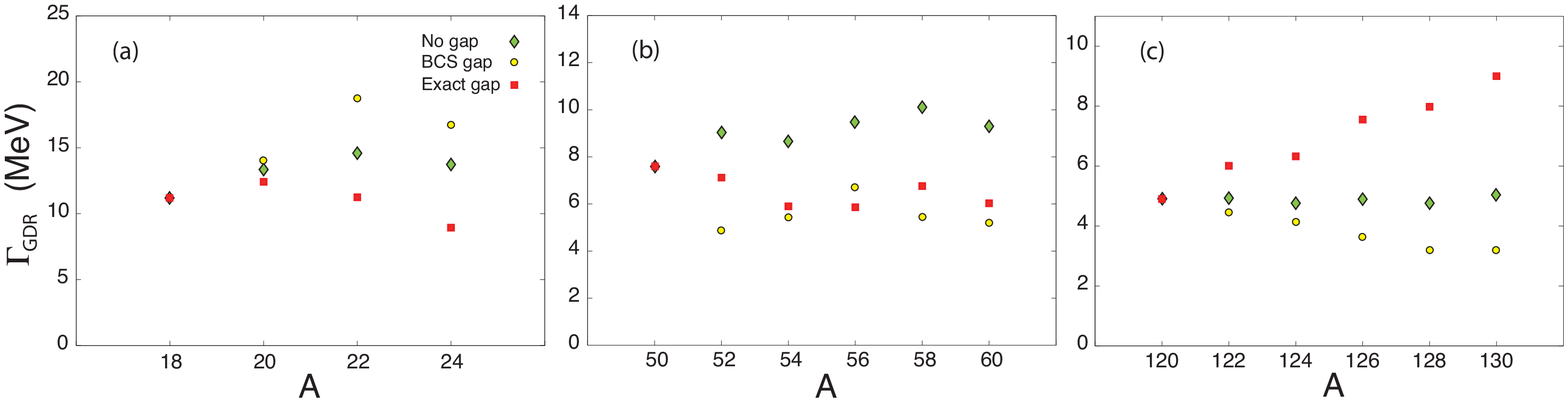}
        \caption{(Color online) GDR width $\Gamma_{GDR}$ obtained by using parameter selection I as a function of mass number A for oxygen (a), calcium (b) and tin (c) isotopes. Notations are as in Fig. \ref{F1}.
        \label{width}}
    \end{figure}

Shown in Figs. \ref{sOF1} -- \ref{sSnF1} are the GDR strength functions $S(\omega)$ obtained within the parameter selection I for oxygen, calcium and tin isotopes, respectively. It is clear from these figures that, by using the same values of $F_1$, which are determined to fit the GDR width in $^{18}$O, $^{50}$Ca, and $^{120}$Sn, for the rest of nuclei in the corresponding isotopic chains, the differences between the predictions obtained without pairing, including BCS pairing, and exact pairing increase with the neutron number, and, therefore, the mass number A. The GDR width obtained from Eq. (\ref{Gamma}) in the calculations without pairing does not change much with A, as shown by the diamonds in Fig. \ref{width}, where $\Gamma_{GDR}$ is between around 11 and 14 MeV for oxygen,  around 8 and 10 MeV for calcium, and around 5 MeV for tin isotopes. The results of calculations including the BCS gap show that the GDR width for oxygen isotopes increases first when A increases from 18 to 22, but drops when A increases further from 22 to 24. For tin isotopes, the BCS results show a GDR width that decreases with increasing A. The predictions obtained including the exact pairing expose an opposite trend, namely the GDR width decreases when A increases from 20 to 24, but always increases with A for tin isotopes. Calcium isotopes correspond to the intermediate stage, where some oscillating behavior is seen in the dependence of the GDR width on the mass number as shown by the results of the calculations including the BCS and exact pairing gaps.

\begin{figure}
       \includegraphics[width=16cm]{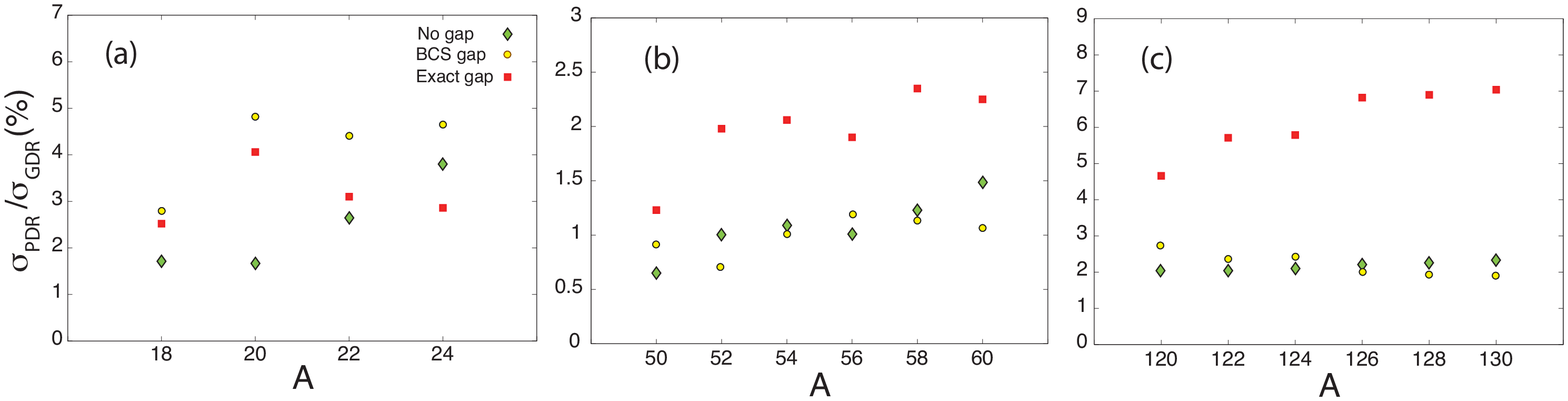}
        \caption{(Color online) The fraction of PDR in the total integrated cross section of the GDR obtained by using parameter selection I as a function of mass number A for oxygen (a), calcium (b) and tin (c) isotopes. Notations are as in Fig. \ref{F1}.
        \label{EWSR_F1}}
    \end{figure}

The most important feature seen in Figs. \ref{sOF1} -- \ref{sSnF1} is the enhancement of the PDR under the effect of pairing. This effect is particular strong when exact pairing is used  for medium and heavy nuclei with larger A, as displayed in Fig. \ref{EWSR_F1}, which shows that the PDR obtained by using exact pairing exhausts from around 4.5 to 7$\%$ of the total GDR integrated cross section for tin isotopes, and it becomes stronger with increasing A. Meanwhile those obtained without pairing and including BCS pairing contribute only around 2 -- 3$\%$ to the total EWSS of GDR. 
\subsubsection{Predictions by using parameter selection II}
\begin{figure}
       \includegraphics[width=16cm]{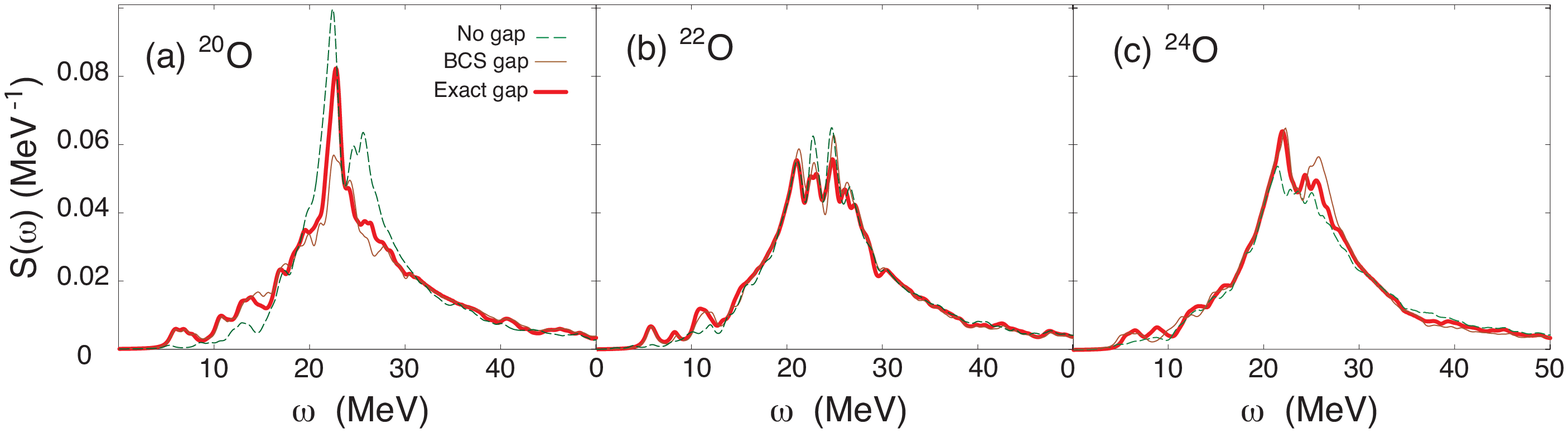}
        \caption{(Color online) GDR strength functions (\ref{S}) for oxygen isotopes obtained by using parameter selection II.  Notations are as in Fig. 
        \ref{sOF1}. 
        \label{sO}}
    \end{figure}

\begin{figure}
       \includegraphics[width=16cm]{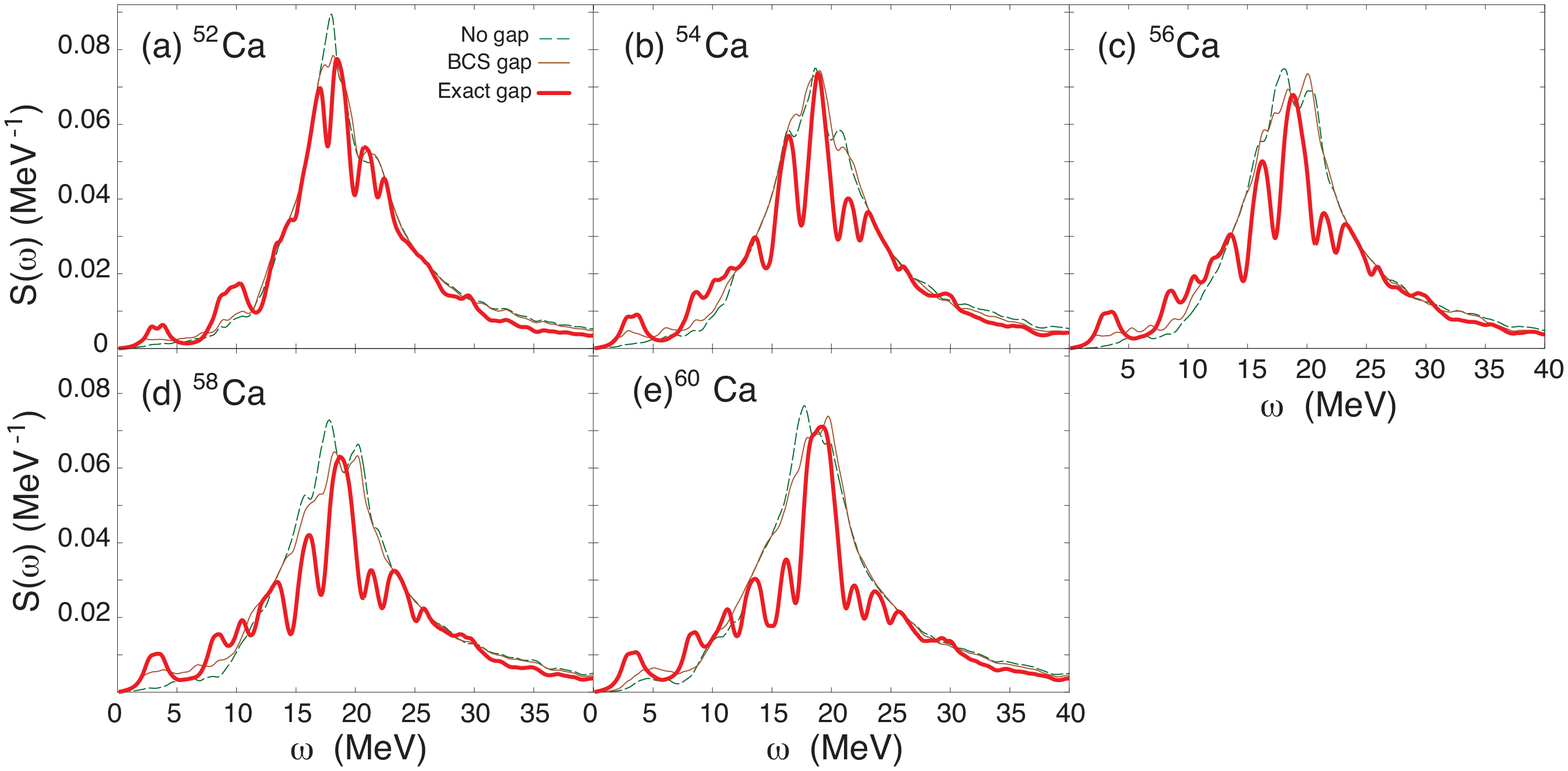} 
       \caption{(Color online) Same as Fig. \ref{sO} for calcium isotopes.
         \label{sCa}}
    \end{figure}

\begin{figure}
       \includegraphics[width=16cm]{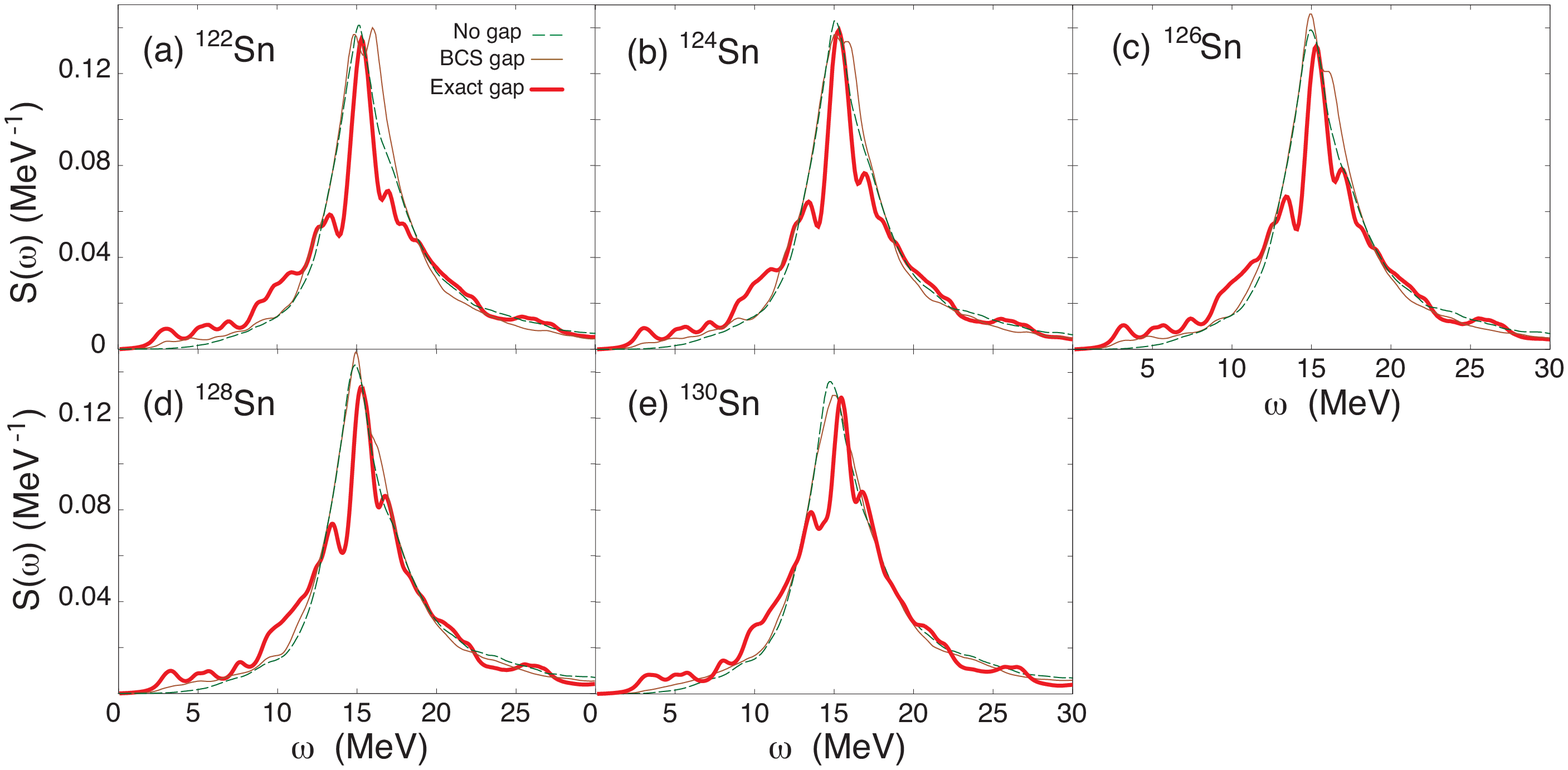}
        \caption{(Color online) Same as Fig. \ref{sO} for tin isotopes.
        \label{sSn}}
    \end{figure}

\begin{figure}
       \includegraphics[width=16cm]{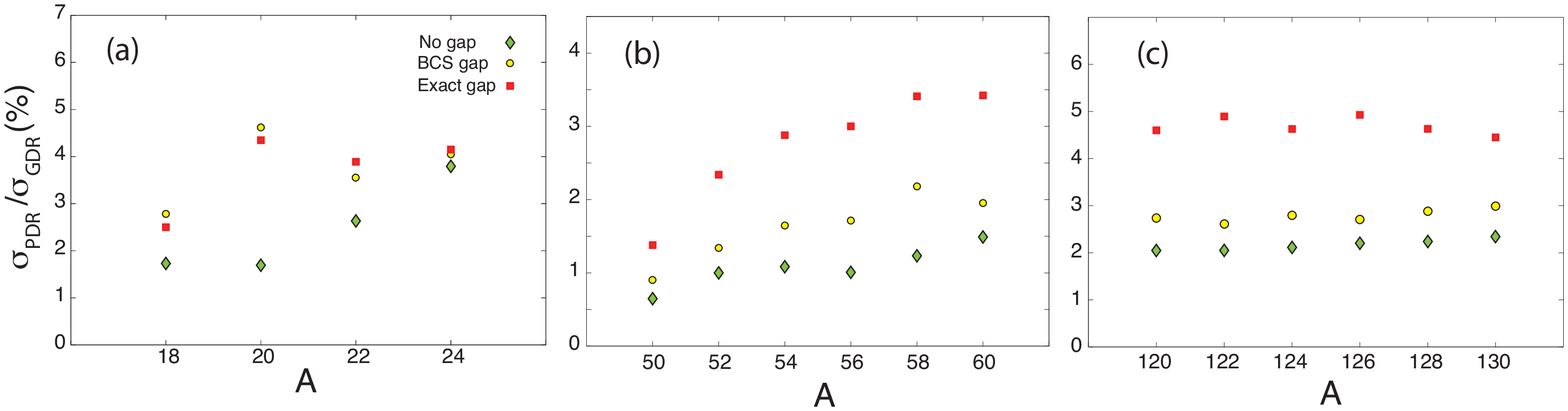}
        \caption{(Color online) The same as in Fig. \ref{EWSR_F1} but obtained by using the parameter selection II.
        \label{EWSR_F2}}
    \end{figure}

The GDR strength functions obtained by using the parameter selection II are shown in Figs. \ref{sO} -- \ref{sSn} for oxygen, calcium, and tin isotopes respectively. As compared to those predicted by using the parameter selection I, the shapes of the GDR obtained including BCS and exact pairing gaps significantly change because they now have the same width as that of the GDR obtained without pairing.  The PDR contribution to the total GDR integrated cross section obtained by using exact pairing becomes slightly larger (around 4$\%$) for $^{22,24}$O as compared to the case when the BCS gap is used (Fig. \ref{EWSR_F2}). Particularly stronger becomes the PDR obtained including pairing for calcium isotopes, where the ratio $\sigma_{PDR}/\sigma_{GDR}$ is between 1.5 and 2$\%$ in the BCS case and between around 2.8 and 3.5$\%$ in the exact pairing case for $^{54-60}$Ca. Meanwhile, within the parameter selection I this ratio does not exceed 2.4$\%$ by using exact pairing. The increase of GDR splitting with A for calcium isotopes is also much stronger than that predicted by using the parameter selection I (Fig. \ref{sCa}). For tin isotopes, pairing obviously enhances the EWSS of the PDR but this enhancement  does not seem to strongly depend on the mass number A. The ratio $\sigma_{PDR}/\sigma_{GDR}$ for tins is between 4.5 and 5$\%$ in the case with exact pairing, and between 2.6 and 3$\%$ when the BCS gap is employed.

It is worth mentioning that, in Ref. \cite{PDM}, the Hartree-Fock method with two versions of Skyrme interaction, SGII and SIII, was employed to generate the single-particle energies. The results obtained for the GDR therein are qualitatively similar to those of the present paper, which are also comparable to the predictions by state-of-the-art microscopic approaches such as those quoted in Fig. 4 of Ref. \cite{PDM} and in Ref. \cite{Terazaki}. 
This gives us the justification and confidence in the description within the PDM in the present paper. 

\begin{figure}
       \includegraphics[width=16cm]{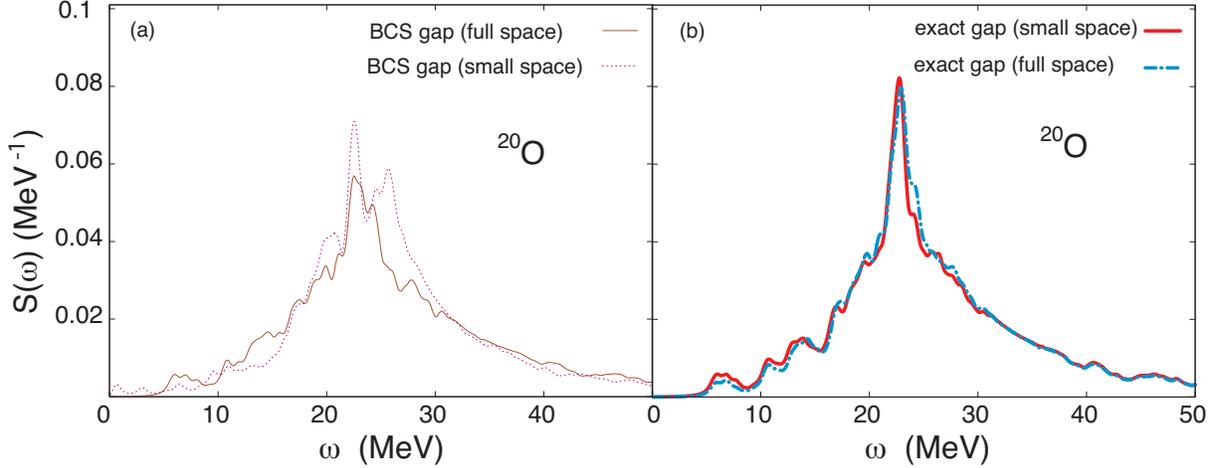}
        \caption{(Color online) $E1$ strength functions of $^{20}$O obtained by using BCS pairing (a) and exact pairing (b). In (a) the (brown) thin solid line is the same as that in Fig. \ref{sO} (a), whereas for the (purple) dotted line the BCS gap was calculated by using the same 7 neutron spherical orbitals employed in calculating the exact pairing gap.
        In (b) the (red) thick solid line is the same as that in Fig. \ref{sO} (a), whereas the (blue) dash-dotted line is obtained by extending the exact pairing gap to the all neutron single-particle orbitals. (See text.)
        \label{Stest}}
    \end{figure}

Last but not least, as has been mentioned earlier in Sec. III A, the BCS gaps used in the calculations of the strength functions (the (brown) thin solid lines in Figs.  
\ref{sOF1} -- \ref{sSnF1}, \ref{sO} - \ref{sSn}) are obtained by solving the BCS equations within the entire neutron single-particle spectra, whereas the exact gaps  are calculated in the limited single-particle spaces because the sizes of matrices to be diagonalized cannot be too large. A question naturally arises on how the use of the same configuration space in both BCS and exact calculations of the pairing gaps affects the comparison. The answer to this question is given in Fig. \ref{Stest}, where the $E1$ strength functions obtained by using the BCS and exact pairing gaps in $^{20}$O are plotted. In Fig. \ref{Stest} (a) the result obtained by using the BCS pairing gap in the entire neutron single-particle spectrum is compared with that obtained by using the BCS calculated in the limited space, the same as that employed in the exact diagonalization ((purple) dotted line). For the latter, the pairing interaction parameter $G$ is reduced by around 24\% to retain the same value of the BCS gap. Because the $u_j$ and $v_j$ coefficients for the levels outside this limited space become 1 (0) and 0 (1), the parameter $F_1$ has to be increased by 92\% so that the width of the strength function remains the same. The comparison clearly shows that the use of the BCS gap obtained in a small configuration space is not able to produce any noticeable PDR structure. Instead the strong quasiparticle-phonon coupling $F_1$ spreads out the low-energy part of the $E1$ strength function into a long tail down to very low energy. The comparison in Fig. \ref{Stest} (b) shows a different picture. Here, to produce the (blue) dash-dotted line, the exact pairing gap obtained by exact diagonalization in the limited configuration space is extended to all the neutron single particle levels outside the set by assuming that they all have the quasiparticle energies given in Eq. (\ref{qpenergy}). The value of $F_1$ is reduced only by 20\% in this case so that the width of the strength function remains unchanged. By comparing Figs. \ref{Stest} (a) and \ref{Stest} (b), one can see that the slight difference between the (red) thick solid and (blue) dash-dotted lines in Fig. \ref{Stest} (b) is a further justification for the preference of using the exact pairing gap to describe the PDR in neutron-rich nuclei. 
\section{Conclusions}
The present paper studied the effect of superfluid pairing on the PDR in light, medium and heavy neutron-rich oxygen, calcium and tin isotopes. Beside the conventional BCS gap, the exact pairing gap obtained by diagonalizing the pairing Hamiltonian is also employed to calculate the strength function of the GDR in these nuclei within the framework of the PDM. Two sets of parameter selection of the PDM are adopted.
In the first set, the values for the parameter of quasiparticle-phonon coupling  are chosen to describe the GDR in stable nuclei $^{18}$O, $^{50}$Ca and $^{120}$Sn, in the cases without pairing, and including BCS or exact pairing. These values are then kept fixed and extended to the calculations for other nuclei in the same isotopic chains. In the second parameter selection, the values for this parameter is readjusted when BCS and exact pairing are included so that the obtained GDR width remains the same as that predicted in the case without pairing. The analysis of the numerical calculations allows us to make the following conclusions:

1) Exact pairing decreases the two-neutron separation energy in light nuclei, but increases it in heavy nuclei as compare to that obtained within the BCS theory;

2)  Exact pairing significantly enhances the PDR in medium (calcium) and heavy (tin) nuclei, whereas the BCS pairing causes a much weaker effect as compared to the case when pairing is neglected. This observation indicates that BCS pairing might not be sufficient to describe the PDR in medium and heavy neutron-rich nuclei; 

3) The significant change in the line shape of the GDR with increasing the mass number A, which takes place within the calculations using parameter selection I, indicates that the values for the model's parameters cannot be kept fixed when the calculations are extended to the nuclei in the vicinity of the neutron drip line.
This includes the parameters of the nuclear mean field such as the parameters of the Woods-Saxon potential or the parameters of effective interactions such as various Skyrme types, which are used in microscopic calculations of the GDR and PDR.

The obtained results may serve as a hint to clarify while several microscopic approaches, mentioned in the Introduction, are in disagreement regarding the strength and fine structure of the PDR. The present paper also emphasizes the necessity of using exact pairing, whenever possible, instead of the BCS one or the HFB average pairing gap in the future study of the PDR. 

The results of this work also demonstrates that, albeit being a semi-microscopic model with phenomenological assumptions for the parameter of GDR coupling to ph configurations, the PDM is able to give reasonable predictions for the GDR as well as PDR, which are qualitatively similar to those offered in the state-of-the-art microscopic approaches. The drawback of the PDM is that it is a simple model with phenomenological parameters; therefore, it does not allow a quantitative study of the structure of the PDR such as its collectivity and transition densities. 
In order to carry out this task, the GDR phonon should be made as a collective superposition of two-quasiparticle states within a microscopic framework such as the QRPA by using some effective interaction such as the Skyrme force. By solving the corresponding set of QRPA equations one can determine the structure of the GDR phonons, namely their energies and the corresponding QRPA $X$ and $Y$ amplitudes. This project is now under consideration and we hope to report the results in a forthcoming study.

\acknowledgments
The numerical calculations were carried out using the {\scriptsize FORTRAN IMSL}
Library by Visual Numerics on the RIKEN Integrated Cluster of Clusters (RICC) system. 


\begin{thebibliography}{99}
\bibitem{P1}J. Chambers {\it et al.}, Phys. Rev. C {\bf 50}, R2671 (1994).
\bibitem{P2}A. Liestenscheneider {\it al.}, Phys. Rev. Lett. {\bf 86}, 5442 (2001).
\bibitem{P3}E. Tryggestad  {\it et al.}, Phys. Lett. B {\bf 541}, 52 (2002).
\bibitem{P4}T. Aumann {\it et al.}, Eur. Phys. J. A {\bf 26}, 441 (2005).
\bibitem{P5}J. Gibelin {\it et al.}, Nucl. Phys. A {\bf 788}, 153 (2007).
\bibitem{P6}O. Wieland and A. Bracco, Prog. Part. Nucl. Phys. {\bf 66}, 374 (2011).
\bibitem{Sagawa}H. Sagawa and T. Suzuki, Phys. Rev. C {\bf 59}, 3116 (1999).
\bibitem{Vretenar}D. Vretenar, N. Paar, P. Ring, and G.A. Lalazissis, Nucl. Phys. A {\bf 692}, 496 (2001); N. Paar, T. N\u{i}ks\'ic, D. Vretenar, and P. Ring,
Phys. Lett. B {\bf 606}, 288 (2005).
\bibitem{Ansari}A. Ansari and P. Ring, Phys. Rev. C {\bf 74}, 054313 (2006). 
\bibitem{Litvinova} E. Litvinova, P. Ring, and D. Vretenar, Phys. Lett. B {\bf 647}, 111 (2007).
\bibitem{Sarchi}D. Sarchi, P.F. Bortignon, and G. Col\`o, Phys. Lett. B {\bf 601}, 27 (2004).
\bibitem{Terazaki}J. Terazaki and J. Engel, Phys. Rev. C {\bf 74}, 044301 (2006), J. Terazaki and J. Engel, Phys. Rev. C {\bf 82}, 034326 (2010).
\bibitem{Dang}N. Dinh Dang, T. Suzuki, and A. Arima, Phys. Rev. C {\bf 61}, 064304 (2000).
\bibitem{PDM} N. Dinh Dang, V. Kim Au, T. Suzuki, and A. Arima, Phys. Rev. C {\bf 63}, 044302 (2001).
\bibitem{Ebata}S. Ebata {\it et al.}, Phys. Rev. C {\bf 82}, 034306 (2010).
\bibitem{Lanza}E. Lanza {\it et al.}, Phys. Rev. C {\bf 84}, 064602 (2011).
\bibitem{Inakura}T. Inakura, T. Nakatsukasa, and K. Yabana, Phys. Rev. C {\bf 84}, 021302 (R) (2011).
\bibitem{Gamba}D. Gambacurta, M. Grasso, and F. Catara, Phys. Rev. C {\bf 84}, 034301 (2011).
\bibitem{Tsoneva}N. Tsoneva and H. Lenske, J. Phys.: Conf. Ser. {\bf 366}, 012043 (2012).
\bibitem{Volya}A. Volya, B. A. Brown, and V. Zelevinsky, Phys. Lett. B {\bf 509}, 37
(2001).
\bibitem{CE}N. Quang Hung and N. Dinh Dang, Phys. Rev. C {\bf 79}, 054328 (2009).
\bibitem{WS}S. Cwiok {\it et al.}, Comput. Phys. Commun. {\bf 46}, 379 (1987).
\bibitem{PDM2}N. Dinh Dang and A. Arima, Phys. Rev. C {\bf 68}, 044303
    (2003).   
\bibitem{Bohr}A. Bohr and B.R. Mottelson, {\it Nuclear Structure} (Benjamin, New York, 1969), Vol. 1, p. 170.
\bibitem{exgap}G. Audi, A.H. Wapstra, and C. Thibault, Nucl. Phys. A {\bf 729}, 337 (2003).
\bibitem{PRC76}N. Quang Hung and N. Dinh Dang, Phys. Rev. C {\bf 76}, 054302 (2007), Ibid. {\bf 77}, 029905(E) (2008).
\bibitem{PDMT}N. Dinh Dang and A. Arima, Phys. Rev. Lett. {\bf 80}, 4145 (1998); N. Dinh Dang and A. Arima, Nucl. Phys. A {\bf 636}, 427 (1998).
 \bibitem{O18}J. Ahrens {\it et al.}, Nucl. Phys. A251, 479 (1975).
\bibitem{Berman}B.L. Berman and S.C. Fultz, Rev. Mod. Phys. {\bf 47}, 713 (1975).
\end{thebibliography}
\end{document}